# Conquering the Rayleigh scattering limit of silica glass fiber at visible wavelengths with a hollow-core fiber approach

Shou-fei Gao,[1] Ying-ying Wang,[1,*] Wei Ding,[2,3] Yi-Feng Hong,[1] and Pu Wang[1,4]

[1] Beijing Engineering Research Center of Laser Technology, Institute of Laser Engineering, Beijing University of Technology, Beijing , China
[2] Guangdong Provincial Key Laboratory of Optical Fiber Sensing and Communications, Institute of Photonics Technology, Jinan University, Guangzhou 510632, China
[3] photonicsweiding@163.com
[4] wangpuemail@bjut.edu.cn
*Corresponding author: dearyingyingwang@hotmail.com

**The ultimate limit on fiber loss is set by the intrinsic Rayleigh scattering of silica glass material. Here, we challenge this limit in the visible region by using a hollow-core fiber approach. Two visible-guiding hollow-core conjoined-tube negative-curvature fibers are successfully fabricated and exhibit the overall losses of 3.8 dB/km at 680 nm and 4.9 dB/km at 558 nm respectively. The loss of the latter fiber surpasses the Rayleigh scattering limit of silica glass fiber in the green spectral region by 2 dB. Numerical simulation indicates that this loss level is still much higher than the fundamental surface scattering loss limit of hollow-core fiber.**

After over 40 years of technological refinement, the Rayleigh scattering loss (RSL), caused by microscopic density fluctuation frozen in bulk glass [1], has become the dominant loss constituent in state-of-the-art silica glass fiber (SGF) throughout most of its operating wavelength. The fact that the RSL is fundamentally insurmountable casts a shadow on future applications of optical fiber ranging from communications to sensing and laser. First, the most efficient light transmission window in a fiber is restricted near 1.55 μm wavelength with the minimum loss of ~ 0.14 dB/km [2] and the bandwidth of some tens of THz [3]. In order to send signals through long distance in a fiber, the carrier wavelength of light usually has to be converted to the telecom band [4], introducing great complexity in the design of long-haul optical fiber systems. Second, the $\lambda^{-4}$ wavelength dependence of the RSL is against the trend of employing advanced fiber laser technology in versatile areas, for instance, biochemical imaging and materials-processing, toward shorter and shorter wavelengths for better resolution [5, 6].

Utilizing fluoride glass or multi-component oxide glass, which is characterized by lower glass transition temperature and thus lower density fluctuation, is expected to reduce the RSL by one order of magnitude, e.g., ~ 0.01 dB/km RSL in fluoride glass at 2.5 μm [7] and ~ 0.05 dB/km RSL in sodium silicate glass at 1.55 μm [8]. However, in realistic fiber drawing, significantly high extrinsic losses from impurity, crystallization, defect, etc, have not yet been overcome [7], leaving these soft glass fibers inferior to SGF in terms of propagation attenuation.

Since their first demonstration [9], the possibility that hollow-core photonic bandgap fibers (HC-PBGFs) can bypass the RSL limit of SGF has been highlighted and eagerly anticipated. HC-PBGF fills its core with dry air whose RSL is more than 200 times lower than in silica glass [10]. The cladding of HC-PBGF is formed by a 2D periodic array of air holes with the purpose of completely eliminating light leakage in all transverse directions [11]. However, it was later identified that there exists another loss mechanism in HC-PBGFs, namely surface scattering loss (SSL). The SSL arises from the surface roughness frozen at the glass-air interfaces due to surface capillary waves [12] and has become the dominant loss component in state-of-the-art HC-PBGFs. Because of its thermodynamic origin, the SSL also sets an insuperable barrier and causes the minimum loss of HC-PBGFs to be at 1.7 dB/km [13]. Furthermore, in the cladding area of HC-PBGF, the presence of very thin but mechanically mandatory glass struts (>50 nm) tends to close the photonic bandgaps (PBGs) of higher orders [14], resulting in only a single PBG usable for light guidance. This makes geometrical downscaling the sole but challenging approach for shorter wavelength operation [15]. Consequently, the wavelength dependence of the minimum loss in HC-PBGF mainly follow the $\lambda^{-3}$ law of the SSL [12], just minor improvement from the $\lambda^{-4}$ law of the RSL.

The hope of beating silica's RSL limit is then passed on to another type of hollow-core fiber (HCF), referred to as hollow-core negative-curvature fiber (HC-NCF) [16]. This type of HCF historically stems from Kagome-structured HCF [17] initially famed for its broadband light guidance [18]. Since the discovery of the important negative-curvature (or hypocycloid) core wall shape [19], great progresses have been made in the design and fabrication of low loss HC-NCF [20-22], pushing the loss of HC-NCF to the same level of HC-PBGF, e.g., 2 dB/km at 1512 nm in the conjoined-tube NCF [23] and 1.3 dB/km at 1450 nm in the nested NCF [24]. Unlike HC-PBGF, the glass walls in HC-NCF only act as anti-resonant (AR) reflecting elements [25] with no other adverse effects [26], which facilitate light guidance at visible or even UV wavelengths in the form of either fundamental [27, 28] or higher-order transmission bands [29]. More importantly, the main loss contribution in state-of-the-art HC-NCFs [23, 24] is still the confinement (or leakage) loss (CL), whose $\lambda^{-1}$ wavelength dependence [30, 16] differs dramatically with that of the RSL.

According to all the above analyses, we attempt to challenge the RSL limit of SGF in the visible region by utilization of the newly-developed conjoined-tube NCF (CTF for short) technique. We realize the first, to the best of our knowledge, visible-guiding optical fibers with the overall loss beneath the RSL limit of SGF. The advantages of higher-order-band light guidance and $\lambda^{-1}$ wavelength scaling law in the CTF are fully exploited. The obtained minimum losses of 3.8 dB/km at 680 nm and 4.9 dB/km at 558 nm represent the lowest reported figures in the visible spectral region among all forms of optical fibers and stride across the RSL limit of SGF [3]. We numerically calculate different loss contributions and point out that the current loss level of CTF is still far apart from the fundamental SSL limit. To further reduce the CL, more elaborate fiber structures, e.g., adding more glass layers in the radial direction [23, 24], would be needed. Our results validate that HCF technique does have the capability of circumventing some intrinsic optical fiber loss limits.

In experiment, two visible-guiding CTFs (the red-guiding fiber *a* and the green-guiding fiber *b*) are fabricated by using a modified stack-and-draw method. Fiber *a* has an air core with the inscribed diameter of 25 µm, six conjoined-tubes (CTs) in the cladding, and a glass sheath with the outer diameter of 230 µm [Fig. 1(a)]. In terms of the CTs, the average glass wall thicknesses of the negative-curvature core-surround ($t_1$), the central bar ($t_2$) and the positive-curvature jacket conjunction ($t_3$) are measured to be 0.84 µm, 0.75 µm and 0.82 µm respectively [Fig. 1(b)]. The average inter-CT gap is ~2.7 µm, and the uniformity of structure is well-maintained along the entire length of the fiber. Loss measurement is carried out using the cutback method from 260 m to 10 m. A supercontinuum source is butt-coupled to the input end of the fiber, which is loosely looped on a table with the radius of 30 cm. The output end of the fiber is monitored by a CCD camera to guarantee single modedness and then connected to an optical spectral analyzer through a magnetic clamp bare fiber adaptor. Multiple cleaves are conducted at fiber's output end to ensure that the recorded transmission spectra are repeatable and have very little variations. Two transmission bands are manifest in Fig. 1(c). In the first band, from 960 nm to 1276 nm, corresponding to the 2nd order AR band of all the glass membranes [25], the minimum loss of 2.7 dB/km appears at 1150 nm [Fig. 1(e)], and a number of loss peaks can be attributed to Fano resonance induced extra losses [31]. The 3rd order AR band, from 653 nm to 706 nm [Fig. 1(d)], exhibits a much smoother loss spectrum with those Fano resonance induced loss peaks fading out (see detailed analysis below). In this red-guiding band, the propagation attenuation is measured to be 3.4-4.6 dB/km, the same level of the RSL of SGF.

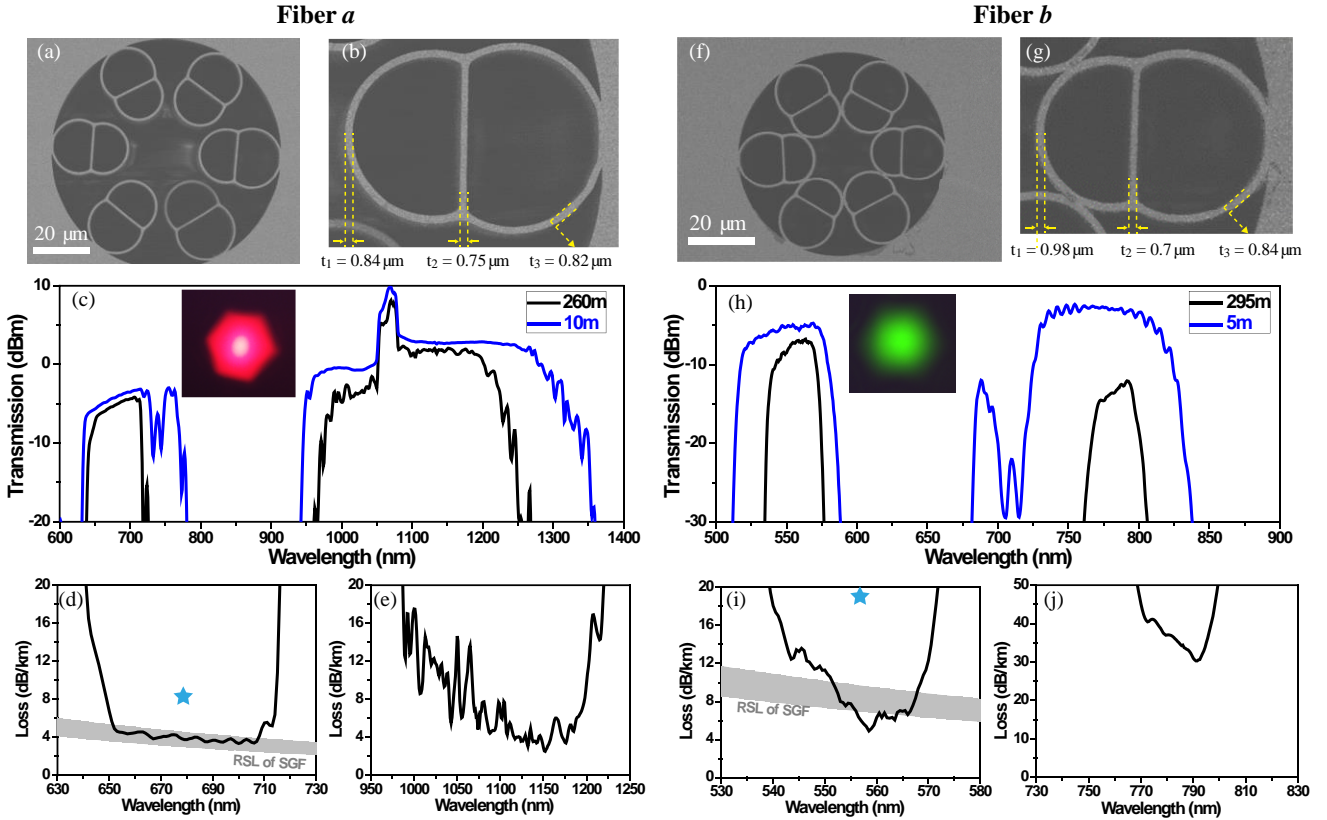

Fig.1. Ultralow loss, visible-guiding CTFs. (a) Scanning electron microscope (SEM) image of the red-guiding fiber *a*. (b) Zoom-in of one CT in fiber a with the three glass wall thicknesses labeled. (c) Measured transmission spectra of fiber a with the lengths to be 10 m (blue curve) and 260 m (black curve) respectively. (d, e) Loss spectra of the 3rd order and 2nd order AR bands respectively. The inset in (c) is a near-field (NF) intensity map at the fiber output after a bandpass filter at 700 nm. (f) SEM image of the green-guiding fiber *b*. (g) Zoom-in of one CT in fiber b. (h) Measured transmission spectra of fiber b with the lengths to be 5 m and 295 m respectively. Inset is the NF pattern at 532 nm. (i, j) Loss spectra of the 3rd/4th order and 2nd/3rd order AR bands respectively. The gray stripes in (d) and (i) outline the RSL range of SGF. The star symbols represents the loss level of commercial SGF.

The challenges in downscaling the dimension of HCF have been well recognized in literature [27-29, 32]. Small sized structures are more susceptible to the variations of fiber drawing parameters, e.g., pressure, temperature and drawing speed, leading to severe deformations. For example, in terms of HC-PBGF, the theoretically predicted minimum loss according to the $\lambda^{-3}$ law of SSL [12] is around 50 dB/km at visible wavelengths, while the fabricated HC-PBGF exhibits loss higher than 800 dB/km [32]. With regard to the HC-NCF having a single ring of tubes, the minimum loss follows well with the $\lambda^{-1}$ law of CL in the infrared region. However, this scaling law alters to between $\lambda^{-2}$ and $\lambda^{-3}$ in the visible due to the increased difficulty in preserving ideal geometry [16]. After a number of trails, we managed to realize a CTF with a just-contacting structure, see Fig. 1(f). We find that the influence from this level of non-ideal geometry of the just-touching CTFs is moderate with respect to the propagation loss, although the influence on the bending loss is more pronounced, as will be discussed later.

As shown in Figs. 1(f, g), the core diameter of fiber *b* has decreased remarkably to 18 µm, while the glass wall thicknesses are 0.98 µm ($t_1$), 0.7 µm ($t_2$) and 0.84 µm ($t_3$), respectively, with $t_2$ being notably different from $t_1$ and $t_3$. As pointed out by the intuitive multi-layered model [33,34], the three glass walls in our CTF can be viewed as a series of cascaded Fabry–Pérot resonators. For the purpose of light guidance, they actually do not need to have the same AR order [35].

Figures 1 (h-j) show the measured transmission and loss spectra of fiber *b*. In the notation of [35], the longer wavelength window can be called as the 2nd/3rd order band, and the shorter wavelength one is the 3rd/4th order band. The minimum losses in the two bands are 30 dB/km at 790 nm and 4.9 dB/km at 558 nm, respectively. The relatively narrow bandwidths (~30 nm) can be attributed to the overlapping of hybrid AR bands and the worse uniformity of the glass wall thickness in fiber *b*.

We plot the RSL of typical SGF in Figs. 1(d, i) with the Rayleigh scattering constant $C_R$ in the range of 0.7–0.9 dB/(km·μm$^4$) [36]. This choice of $C_R$ corresponds to the loss of state-of-the-art SGF of 0.12 - 0.15 dB/km at 1550 nm. It is seen that our CTF *a* has reached the same level of the RSL in the 653 - 706 nm wavelength region and the CTF *b* far surpasses the RSL limit in the 555 - 564 nm wavelength region. At 558 nm, the loss of fiber *b* (4.9 dB/km) is 2 dB lower than the RSL limit of SGF (using $C_R$ = 0.7), indicating that fiber *b* undoubtedly breaks a long-sustained optical fiber loss limit set by silica glass material. It should be also noted that commercial SGFs typically have higher attenuations than the above figures, e.g., 8-10 dB/km at 680 nm and 20-30 dB/km at 558 nm (labeled as star symbols in Fig.1) [37], which are 2-5 times higher than our CTFs. Considering other HCFs, the loss of HC-PBGF at visible wavelengths is usually greater than 500 dB/km [32]. For single ring HC-NCF with core diameter of 26 μm, a loss of 80 dB/km at 532 nm was featured [27]. Recently, a result of 13.9 dB/km at 539 nm has been reported with a very large core diameter of 42 μm [38], which gives rise to severe bending sensitivity.

To elucidate the significances of different loss contributions (CL, SSL, microbending loss-MIBL, and macrobending loss-MABL), we use finite element method (FEM, COMSOL Multiphysics) and high resolution SEM images to calculate the imaginary part of mode effective index for the CL, the mode field distribution for the SSL [12], and the inter-modal power coupling efficiencies for the MIBL [39]. According to the experimental condition, we set the bending radius of the fiber to be 30 cm to take into account the MABL in FEM simulation. As shown in Fig. 2(b), the CL overwhelms in the 2nd order AR band. The characteristic loss peaks in the experimentally measured spectrum are reproduced and their Fano resonance origin is validated by the presence of standing waves along the glass membranes. On the other hand, with the increase of the core diameter to wavelength ratio D/λ from 23 to 37, in the 3rd order AR band [Fig. 2(a)], the CL, together with those Fano resonance induced loss peaks, is substantially suppressed and becomes the secondary loss contribution.

To evaluate the MIBL caused by random microbends along the fiber, we hypothesize that the power spectral density of the fiber curvature follows an inverse power law of C(Δβ) = $C_0$/Δβ$^2$, with Δβ the spatial frequency in the longitudinal direction [39]. By choosing $C_0$ = 1/0.35, we obtain a good agreement between the simulated and the measured loss spectra across both two transmission bands for fiber *a*. In the 3rd order AR band, the MIBL becomes the dominant loss, implying that enhancing straightness of CTF by more dedicated polymer coating technique (e.g., "wet-on-dry" double coating) will be of great importance in future refinement. At the time of this experiment we use a single layer coating with a relatively big segment modulus (35 MPa). It is suggested that when the segment modulus decrease to for instance 1MPa (as applied in the primary coating layer of the microbending-insensitive SGF [40]), the tension force frozen in a fiber can be effectively released, and the fiber straightness can be greatly improved. This MIBL issue needs to be studied in more detail and could be ameliorated substantially given its extrinsic attributes.

According to the simulated spectra in Figs. 2(a, b), we estimate the SSL to be nearly one order of magnitude lower than the overall loss, leaving plenty of room for further improvement by more sophisticated CTF structures. Besides the SSL, the RSL (not shown here), which is another intrinsic loss source in CTF, is well below 0.1 dB/km thanks to the <0.02% mode field overlap with glass.

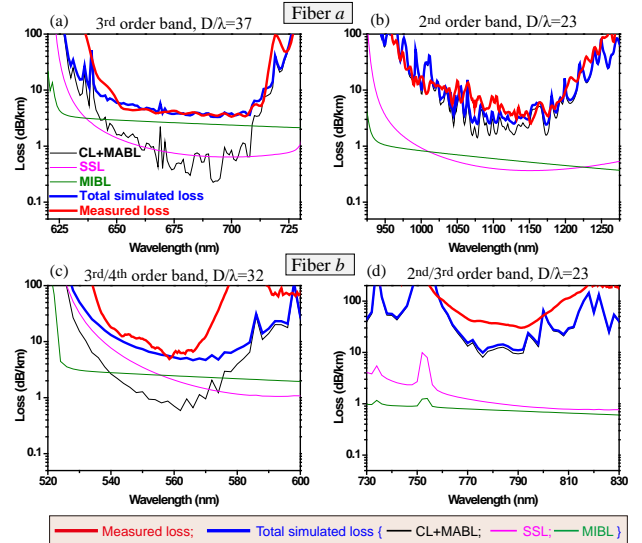

Fig. 2. Comparison of experimental and simulated loss spectra in (a) the 3rd order and (b) the 2rd order bands of fiber *a*. The same in (c) the 3rd/4th order and (d) the 2nd/3rd order bands of fiber *b*.

All the above analyses are also applicable for fiber *b*, as shown in Figs. 2(c, d). The less consistency between experiment and simulation may be due to the worse structural uniformity in the longitudinal direction, which has been observed in fiber *b*.

For practical employment of above visible-guiding CTFs, the MABL property need to be characterized. Figure 3 shows the measured MABL spectra with the bending radii *R* to be 13 cm, 10 cm, and 7.5 cm, respectively. For fiber *a*, the MABLs at 670 nm are 3.5 dB/km, 19 dB/km, and 50 dB/km, respectively. For fiber *b*, in its green-guiding band, the MABLs become even higher with the figures to be 20 dB/km, 28 dB/km, and 90 dB/km, respectively, probably due to the relatively worse structural uniformity along the fiber length. At small bending radii, the MABL spectra of fiber *b* show some oscillation features, indicating that some neighboring CTs have contacted tightly with one another and result in some Fano resonance induced loss peaks. It is noteworthy that the D/λ ratios of 32 and 37 in Fig. 3 are in the same level of many recently-reported Kagome fibers and single-ring-tube NCFs [16,28], where a great number of useful applications have already been exploited. When the D/λ decreases, for example entering into the longer wavelength bands of fibers *a* and *b*, the MABLs become negligible, in good agreement with the results in [23].

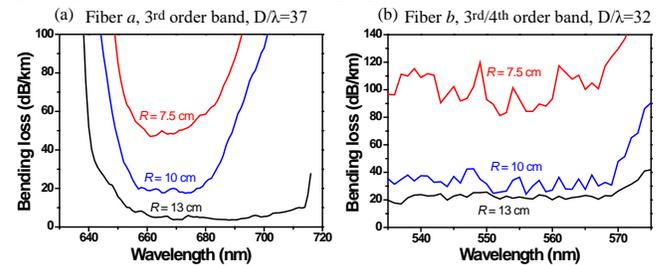

Fig. 3. Measured MABLs of (a) fiber *a* in the red-guiding band and (b) fiber *b* in the green-guiding band. In experiment, more than 50 turns of fiber coils are conducted.

In conclusion, since SGF cannot exceed the fundamental RSL limit and soft glass fibers have not fulfilled low RSL owing to the material impurity induced extrinsic losses, HCF undertakes the task of extending the range of possibilities in optical fibers. However, the first generation of HCF, i.e., HC-PBGF, is influenced by the detrimental glass struts, resulting in the closure of higher order PBGs and the overwhelming SSL with a $\lambda^{-3}$ wavelength scaling law. Fortunately, HC-NCF eliminates both these two drawbacks and extends the potential of HCF technique. In this work, by leveraging the higher-order-band light guidance and the $\lambda^{-1}$ wavelength dependence of the CL, along with employing our celebrated CTF geometry, we demonstrate an optical fiber loss beneath the RSL limit of SGF by 2 dB for the first time. Apart from the fundamental interest of conquering the RSL limit of SGF, an ultralow loss visible-guiding fiber also gains great momentums from the application aspect.